# On Productions of Net-baryons in Central Au-Au Collisions at RHIC Energies

**Ya-Hui Chen, Guo-Xing Zhang, and Fu-Hu Liu**
*Institute of Theoretical Physics, Shanxi University, Taiyuan, Shanxi 030006, China*
Correspondence should be to addressed to Fu-Hu Liu; fuhuliu@163.com

**Abstract**:
The transverse momentum and rapidity distributions of net-baryons (baryons minus anti-baryons) produced in central gold-gold (Au-Au) collisions at 62.4 and 200 GeV are analyzed in the framework of a multisource thermal model. Each source in the model is described by the Tsallis statistics to extract the effective temperature and entropy index from the transverse momentum distribution. The two parameters are used as input to describe the rapidity distribution and to extract the rapidity shift and contribution ratio. Then, the four types of parameters are used to structure some scatter plots of the considered particles in some three-dimensional (3D) spaces at the stage of kinetic freeze-out, which are expected to show different characteristics for different particles and processes. The related methodology can be used in the analyzes of particle production and event holography, which are useful for us to better understand the interacting mechanisms.

**Keywords**: Transverse momentum distribution, Rapidity distribution, Net-baryons, Au-Au collisions, Tsallis statistics, Scatter plots

## 1. Introduction

In high energy nucleus-nucleus collisions, due to the fast speed and the great mass of the projectile and target nuclei, the collisions can form new substances such as pions, kaons, antiprotons, and so forth which are not obviously exist in the nuclei [1, 2]. In the area filled with new and rich phenomena, there are many issues which are difficult to deal with, and the collisions still stay at the phenomenological stage in understanding [3]. To restore the original process of the collisions by analyzing various properties of final-state particles is a major aspect in high energy nuclear physics [4].

The relativistic heavy ion collider (RHIC) located in the United States and the large hadron collider (LHC) located in Switzerland are the main sources of experimental data of heavy-ion collisions [5–7]. They play very significant roles in the studies of high energy nucleus-nucleus collisions. The transverse momentum and rapidity distributions of different final-state particles can be measured from the experiments. The two distributions are related to the excitation degree of the interacting system and the penetrating power of the projectile nucleus respectively. From the two distributions, we can extract some other distributions and correlations. Scientists worked in fields of high energy and nuclear physics are very interested in experiments performed in the colliders.

Since 1980's or earlier days, many phenomenological models have been proposed to explain the large number of experimental data in high energy particle-particle, particle-nucleus, and nucleus-nucleus collisions [4]. In these models, except for the standard (Boltzmann, Fermi-Dirac, and Bose-Einstein) and other distributions, the Tsallis statistics can describe the thermal source (fireball) and transverse momentum distributions [8–10]. An effective temperature parameter which reflects the excitation degree of interacting system and an entropy index (non-extensive parameter) which describes the degree of non-equilibrium of interacting system are used in the Tsallis statistics. Particularly, the Tsallis statistics describes the temperature fluctuations in two- or three-standard distributions.

Among abundant experimental data, the transverse momentum and (pseudo)rapidity distributions of identified particles are very important. From the two distributions, we can extract some parameters such as the (effective) temperature, entropy index, rapidity shift, and contribution ratio in the framework of a multisource thermal model [11–14]. These parameters can be then used to extract some scatter plots of the considered particles in some three-dimensional (3D) spaces. These scatter plots are expected to show different characteristics for different particles and processes. We then propose a new method for studies of particle production and event holography. In addition, baryon transport is a fundamental aspect of high energy nucleus-nucleus collisions as it pertains to the initial scatterings where the Quark-Gluon Plasma (QGP) is formed. It is a process that is not very well understood and so further theoretical studies are needed.

In this paper, in the framework of a multisource thermal model [11–14] in which each source is described by the Tsallis (transverse) momentum distribution [8–10, 15, 16], we analyze the transverse momentum and rapidity distributions of net-baryons (baryons minus anti-baryons) produced in central gold-gold (Au-Au) collisions at the center-of-mass energies of 62.4 GeV and 200 GeV. Based on the descriptions of the two distributions, some scatter plots of the considered particles in some 3D spaces are structured to show partly event picture in whole phase space. A methodology related to particle production and event holography is presented to better understand the interacting mechanisms.

## 2. The model and method

According to the multisource thermal model [11–14], in high energy nucleus-nucleus collisions, the target nucleus and the projectile nucleus penetrate each other. We can assume that many emission sources are formed in the interacting system. In a given frame of reference such as the center-of-mass reference frame, these sources with different rapidity shifts $y_x$ in the rapidity space can be divided into four parts: a leading target nucleon cylinder (LT) in rapidity interval $[y_{LT\min}, y_{LT\max}]$, a leading projectile nucleon cylinder (LP) in rapidity interval $[y_{LP\min}, y_{LP\max}]$, a target cylinder (TC) in rapidity interval $[y_{T\min}, 0]$, and a projectile cylinder (PC) in rapidity interval $[0, y_{P\max}]$. It is expected that the LT and TC are mainly contributed by the target nucleus, and the LP and PC are mainly contributed by the projectile nucleus. Because the TC and PC are linked at mid-rapidity, they can be regarded as a whole central cylinder. For symmetric collisions such as Au-Au collisions which are studied in the present work, the equations $y_{T\min} = -y_{P\max}$, $y_{LT\max} = -y_{LP\min}$, and $y_{LT\min} = -y_{LP\max}$ can be used. The free rapidity shifts are then $y_{P\max}$, $y_{LP\min}$, and $y_{LP\max}$.

In the rest frame of each source, the source is assumed to emit final-state particles isotropically, and the rapidity of a considered particle is $y'$. Then, the rapidity of the considered particle in the center-of-mass reference frame is

$$y = y_x + y'. \quad (1)$$

According to different origins of different final-state particles, $y_x$ can be in $[y_{LT\min}, y_{LT\max}]$, $[y_{LP\min}, y_{LP\max}]$, $[y_{T\min}, 0]$, or $[0, y_{P\max}]$ corresponding to the LT, LP, TC, and PC, respectively. Generally, $y'$ can be obtained by the definition of rapidity, and $y_x$ can be obtained due to different contribution ratios of the four sources.

In the Tsallis statistics [8–10, 15, 16], let $T$ denote the temperature parameter of the considered source, $q$ denote the non-extensive parameter (entropy index), $N$ denote the number of particles, and $p'$, $m_0$, and $\mu$ denote the momentum, rest mass, and chemical potential of the considered particle, respectively. We have the final-state particle momentum distribution in the rest frame of the considered source to be

$$f(p') = \frac{1}{N}\frac{dN}{dp'} = Cp'^2 \left[1 + \frac{q-1}{T}\left(\sqrt{p'^2 + m_0^2} - \mu\right)\right]^{-\frac{q}{q-1}}, \quad (2)$$

where $C$ is the normalization constant which gives Eq. (2) to be a probability distribution. Generally, the chemical potential $\mu$ can be neglected due to its small value at high energy. Then, Eq. (2) can be simplified as

$$f(p') = \frac{1}{N}\frac{dN}{dp'} = Cp'^2 \left[1 + \frac{q-1}{T}\sqrt{p'^2 + m_0^2}\right]^{-\frac{q}{q-1}}. \quad (3)$$

In the Monte Carlo method, we can obtain the momentum $p'$, the energy $E' = \sqrt{p'^2 + m_0^2}$, the polar angle $\theta'$, the azimuth $\phi' = \phi$, the transverse momentum $p'_T = p'\sin\theta' = p_T$, the $x$-component of momentum $p'_x = p_T\cos\phi = p_x$, the $y$-component of momentum $p'_y = p_T\sin\phi = p_y$, the longitudinal momentum $p'_z = p'\cos\theta'$, the rapidity $y' = 0.5\ln[(E' + p'_z)/(E' - p'_z)]$, the rapidity shift $y_x$, the energy $E = \sqrt{p_T^2 + m_0^2}\cosh y$, the longitudinal momentum $p_z = \sqrt{p_T^2 + m_0^2}\sinh y$, the transverse velocity $\beta_T = p_T/E$, the $x$-component of the velocity $\beta_x = \beta_T\cos\phi$, the $y$-component of the velocity $\beta_y = \beta_T\sin\phi$, the longitudinal velocity $\beta_z = p_z/E$, and so forth, where the quantities with no mark denote those in the center-of-mass reference frame.

By using the above method, we can get many distributions and scatter plots for a given collision process. Generally, $T$ and $q$ are sensitive to the transverse momentum distribution and insensitive to the rapidity distribution. We can use the transverse momentum distribution to determine values of $T$ and $q$. The rapidity shifts and contribution ratios of different sources are usually determined by the rapidity distribution. The contributions of the TC and PC are mainly in the central rapidity region, and the contributions of the LT and LP are mainly in the backward and forward rapidity regions respectively, where the later two regions are also called the fragmentation regions. In the present work, to describe the rapidity distribution as accurately as possible, we may use $T$ and $q$ obtained from $p_T$ distribution at $y = 0$ for the TC and PC, and $T$ and $q$ obtained from $p_T$ distribution at $y = 3.0$

(or 2.9) for the LT and LP.

Particularly, for the considered Au-Au collisions at 62.4 and 200 GeV in the present work, we can describe $p_T$ distribution obtained by the above Monte Carlo method by the analytic expression [10]

$$f_{p_T}(p_T) = \frac{1}{N}\frac{dN}{dp_T} = C_T p_T \sqrt{p_T^2 + m_0^2} \int_{y_{\min}}^{y_{\max}} \cosh y \left[1 + \frac{q-1}{T}\sqrt{p_T^2 + m_0^2}\cosh y\right]^{-\frac{q}{q-1}} dy, \quad (4)$$

where $y_{\min}$ and $y_{\max}$ denote the minimum and maximum rapidities respectively, and $C_T$ is the normalization constant which gives Eq. (4) to be a (normalized) probability distribution. At the same time, $p_x$ ($p_y$) distribution obtained by the above Monte Carlo method can be approximately parameterized to

$$f_{p_{x,y}}(p_{x,y}) = \frac{1}{N}\frac{dN}{dp_{x,y}} = C_{x,y}\left(p_{x,y}^2 + m_0^2\right)\int_{y_{\min}}^{y_{\max}} \cosh y \left[1 + \frac{q-1}{T}\sqrt{p_{x,y}^2 + m_0^2}\cosh y\right]^{-\frac{q}{q-1}} dy, \quad (5)$$

where $C_{x,y}$ is the normalization constant which gives Eq. (5) to be a probability distribution. In the present calculation, we can fit a given $p_T$ distribution by using the Monte Carlo method or analytic expression to obtain values of free parameters $T$ and $q$. It should be noted that the given central $y$ in Eqs. (4) and (5) should be shifted to 0 when we extract $T$ and $q$ because the effect of vectored longitudinal motion have to be subtracted.

## 3. Comparisons and extractions

Figure 1 shows the transverse momentum distributions of net-baryons produced in central Au-Au collisions at (a)–(d) 62.4 and (e)–(h) 200 GeV. From Figures 1(a)–1(d), the corresponding rapidities are $y = 0$, 0.65, 2.3, and 3.0, respectively. From Figures 1(e)–1(h), the corresponding rapidities are $y = 0$, 0.9, 1.9, and 2.9, respectively. The symbols represent the experimental data of the BRAHMS Collaboration [17–19] measured at the RHIC and collected together in [20] in which the longitudinal axis shouldn't be $d^2N/dyd^2p_T$, but $(2\pi p_T)^{-1}d^2N/dydp_T$. The solid curves are the Tsallis fitting results by us. The values of free parameters $T$ and $q$ obtained by using the analytic expression, normalization constant $N_0$ obtained by fitting the experimental data, and $\chi^2/ndf$ (number of degree of freedom) obtained in the fit are given in Table 1, where the last three points in Figure 1(h) have not been included in the calculation of $\chi^2/ndf$ due to the low statistics.

From Figure 1 and Table 1, one can see that $T$ decreases obviously with increase of $y$, and $q$ has a constant value with $y$. This reflects non-equivalent excitations of the sources of net-baryons with different rapidities. The net-baryons with low rapidity are produced in the sources located in the central rapidity region, which are in fact in the TC/PC, for which the degree of excitation is high and the longitudinal speed is low. The net-baryons with high rapidity are produced in the sources located in the fragmentation regions, which are in fact in the LT/LP, for which the degree of excitation is low and the longitudinal speed is high. The fact that the effective temperature decreases with increasing of rapidity renders also that the velocity of radial flow in the fragmentation region is less than that in the central region. In the same region, comparing with 62.4 GeV, the interacting system at 200 GeV has higher excitation degree and larger radial flow. In all rapidity region, both the interacting systems at 62.4 and 200 GeV are close to equilibrium state due to small values of $q$.

Figure 2 presents the rapidity distributions of net-baryons produced in central Au-Au collisions at (a) 62.4 and (b) 200 GeV. The solid symbols represent the experimental data of the BRAHMS Collaboration [17–19] and collected together in [20], and the open symbols are reflected from the data around midrapidity, where ref. [19] is a conference proceeding which means that most points in Figure 2(b) right at $|y|>1$ are preliminary only. The solid curves are our fitting results. From left to right, the dotted or dashed curves presented in each panel are orderly the contributions of the LT, TC, PC, and LP. The maximum values of the net-baryons calculated by us are around the beam rapidities (4.2 at 62.4 GeV and 5.4 at 200 GeV) shown in the figure by the arrows. A few net-baryons extend beyond the beam rapidities due to the forward excitation of nucleons in the beam nuclei and the statistical fluctuations in the Monte Carlo calculation. Other particles have probabilities to extend beyond the beam rapidity [21]. In the calculation, we have used $T$ and $q$ obtained from $p_T$ distribution at $y=0$ for the TC/PC, and $T$ and $q$ obtained from $p_T$ distribution at $y=3.0$ (or 2.9) for the LT/LP. Other free parameters such as $y_{P\max}$, $y_{LP\min}$, $y_{LP\max}$, and the contribution ratio $k$ of LP are shown in Table 2 with the normalization constant $N_0$ and $\chi^2/ndf$. By comparing the modeling results with the experimental data, we find that the experimental rapidity distributions of net-baryons can be well described by the model. Both the rapidity spans $y_{P\max}-0$ and $y_{LP\max}-y_{LP\min}$ in central Au-Au collisions at 200 GeV are greater than those at 62.4 GeV. These trends are natural results.

We would like to point out that the normalization constant for Figure 2 should be known (it should be the number of participant baryons $N_{part}$), but not a free parameter due to baryon number conservation. In fact, the situation is complex, and we have to regard it as a free parameter. As we know, except for the limitation of experimental acceptance range, some baryons in the participant region are not real participants due to the limited scattering cross-section. These no-participants are in fact the spectator baryons. In addition, some baryons in 0–5% and 0–10% centralities discussed in the present work are real spectator baryons. These spectator baryons move straight forward along beam direction and should be excluded in the analysis. Because we don't know exactly the number of spectator baryons, we have to regard the normalization constant as a free parameter. Generally, $N_0 < N_{part} = 357 \pm 8$ [18]. Table 2 confirms this situation.

The scatter plots of 1000 net-baryons in 3D space $p_x$-$p_y$-$y$ in central Au-Au collisions at (a) 62.4 and (b) 200 GeV are given in Figure 3. The black, magenta, olive, and blue balls represent orderly the contributions of LT, TC, PC, and LP from low to high in the rapidity space due to different contribution ratios. Similar to Figure 3, the scatter plots of 1000 net-baryons in 3D (momentum) space $p_x$-$p_y$-$p_z$ and in 3D (velocity) space $\beta_x$-$\beta_y$-$\beta_z$ in central Au-Au collisions at (a) 62.4 and (b) 200 GeV are shown in Figures 4 and 5 respectively. The meanings of balls with different colors are the same as those in Figure 3.

One can see different scatter regions of the four sources at the stage of kinetic freeze-out. The neighbor scatter regions overlap partly in $y$ direction in 3D space $p_x$-$p_y$-$y$, in $p_z$ direction in 3D space $p_x$-$p_y$-$p_z$, and in $\beta_z$ direction in 3D space

$\beta_x$-$\beta_y$-$\beta_z$. The distribution range of $p_z$ is much larger than those of $p_x$ and $p_y$. The distribution pattern in $\beta_z$ direction is very different from those in $\beta_x$ and $\beta_y$ directions. The particle scatter plots are bumpy cylinders in 3D spaces $p_x$-$p_y$-$y$ and $p_x$-$p_y$-$p_z$. The contributions of TC/PC are mainly in the regions with small $|y|$ and $|p_z|$, while the contributions of LT/LP are mainly in the regions with large $|y|$ and $|p_z|$. The density in small $|p_x|$ and $|p_y|$ region is greater than those in other regions. Meanwhile, the particle scatter plot is a bumpy sphere in the velocity space. The densities close to the maximum $|\beta_z|$ are greater than those in other regions. Particularly, in 3D spaces $p_x$-$p_y$-$y$ and $p_x$-$p_y$-$p_z$, the higher center-of-mass energy is, the broader scatter plots are; and in the velocity space, the higher center-of-mass energy is, the higher densities close to the maximum $|\beta_z|$ are.

## 4. Discussions

In the central rapidity region, baryons are produced together with anti-baryons and the net-baryon number is small, where these baryons are produced mainly from gluons and sea quarks. We can use the TC/PC to describe baryon productions in the central rapidity region. In the backward/forward rapidity regions, there are almost only baryons and no anti-baryons, where these baryons are produced from the valence quarks of the target/projectile nuclei. We can use the LT/LP to describe baryon productions in the backward/forward regions. The effective temperature parameter extracted from the Tsallis statistics decreases with increasing the rapidity, which renders that the interactions of gluons and sea quarks in the central rapidity region lose more energy comparing with the interactions of valence quarks in the forward rapidity region. The lost energy is expected to transform mainly to the energy of thermal motions. The more the energy loses, the higher the temperature is.

We would like to point out that the discussion of transverse momentum distributions in Figure 1 doesn't address radial flow. It means that we have obtained the effective temperature, but not the source temperature [22]. It is known that radial flow plays a huge role for transverse momentum distributions, especially for heavy hadrons such as protons. Normally to differentiate between the source temperature and radial flow one does a combined blast wave fit of pions, kaons, and protons. The source temperature describes the thermal motion, and the radial flow describes the blast wave effect. Although the present work doesn't address particularly the radial flow, the extracted parameter $T$ which we call effective temperature contains together the thermal motion and radial flow (blast wave) effect. This means that the succeeding analyzes (Figures 2-5) have not been affected, whether we address particularly the radial flow or not. The only regretful thing is that we don't know the source temperature and radial flow velocity respectively.

In our model, to extract the source temperature $T_0$ [22] and average transverse velocity $\langle u_T \rangle$ of radial flow, we can extract the effective temperatures $T$ and mean transverse momentums $\langle p_T \rangle$ from transverse momentum distributions of pions, kaons, protons, and other light flavor hadrons respectively. On the planes $T - m_0\bar{\gamma}$ (or $T - m_0$ [22]) and $\langle p_T \rangle - m_0\bar{\gamma}$, reasonable linear fittings should give $T = T_0 + am_0\bar{\gamma}$ (or $T = T_0 + a'm_0$) and $\langle p_T \rangle = \langle p_T \rangle_0 + \langle u_T \rangle m_0\bar{\gamma}$, where $\bar{\gamma}$ is the mean Lorentz factor for a given particle, $a$ (or $a'$) is a coefficient related to $\langle u_T \rangle$, $\langle p_T \rangle_0$ is the mean transverse momentum for massless particle, and $T_0$ and $\langle u_T \rangle$ can be obtained by us from the linear relations. It is obviously that we need the transverse momentum distributions for more than three types of particles to do the linear fittings. It is beyond the focus of the present work.

The entropy index shows the same value from the central to forward rapidity regions. This renders that the non-equilibrium degrees for different sources of net-baryons in the interacting system are nearly the same. Although the central rapidity region corresponds to higher source temperature and the forward rapidity region corresponds to higher baryon density, the central rapidity region does not show more advantages or disadvantages comparing with the forward rapidity region in the aspect of non-equilibrium degree for the productions of net-baryons. The higher temperature in the central rapidity region is in fact affected by produced particles which carry more thermal motion energies, and the higher baryon density in the forward rapidity region is caused by the leading nucleons which contribute directly to the baryons. Because the values of entropy index are small, the interacting system is close to the equilibrium state.

In Figures 3 and 4, we have shown $p_x$ and $p_y$ separately. Although the data we used only have $p_T$, it doesn't naturally mean that $p_x$ and $p_y$ have the same distribution. In fact, the anisotropic flow is generally covered in $p_T$ distribution. If we consider the anisotropic flow such as the elliptic flow, $p_x$ and $p_y$ will have different distributions. To extract the anisotropic flow, we need azimuthal information. It is regretful that we neglected this issue in the present work for the purpose of convenience. Although we neglected the anisotropic flow, we still separate $p_x$ and $p_y$ and make 3D plots instead of 2D plots for presenting the methodology. Generally, $p_x$ distribution can be measured in the direction in the reaction plane, and $p_y$ distribution can be measured in the direction out of the reaction plane. In the case of considering the anisotropic flow, $p_x$ and $p_y$ distributions will have different values of parameters. This also explains the reason that the 3D plots in momentum space are different from the actual measurements, $d^2N/dydp_T$, that are reported in experiments in 3D momentum space, and the anisotropic flow is covered in them.

In Figure 5, we have shown a bumpy sphere in the velocity space for presenting partly the interacting event in a whole space image. It seems that there is symmetry between transverse and longitudinal directions. In fact, there is no symmetry due to the density distributions of scatter points in the two directions being totally different. Two small high density areas appear in the regions of large $|\beta_z|$ and small $|\beta_{x,y}|$, and a large low density area appears in the region of small $|\beta_z|$ and various $|\beta_{x,y}|$. It is normal that a few scatter points have large $|\beta_{x,y}|$ at RHIC energies due to some particles in Figure 1 having large $p_T$ (~2 GeV/c). These large $p_T$ result in $\beta_T$ to be around $0.9c$ at $y=0$. Early Lund string model and experimental data gave a $p_T$ of around a few hundred MeV/c in hadronic collisions at low and intermediate energies [23], which results in $\beta_T$ to be around $0.5c$ if $p_T = 600$ MeV/c and $y=0$.

Although Lambda decays affect the rapidity distribution somewhere, we have analyzed the net-baryons but not the net-protons measured by the BRAHMS Collaboration [17–19] due to more abundant data of net-baryons being collected in [20]. It is more convenient for us to use [20] instead of [17–19]. In addition, the typical motivation to study net-baryons is that one expects it to be transported in the initial collisions and there one uses that it is conserved to locate it in the final state. This is why the study of net-baryons is more interesting. Generally, by using the method used in the present work, we can analyze scatter plots for different particles. As a premise, the transverse momentum and rapidity distributions have to be known. If identified particles such as $\pi^+$, $\pi^-$, $K^+$, $K^-$, $p$, $\bar{p}$, and others produced in collisions at different center-of-mass energies are contented the premise, we can structure the corresponding scatter plots for them respectively. Then, we can compare these scatter plots and related root-mean-square quantities carefully. It is expected that some dependences of scatter plots and root-mean-square quantities on particle mass and center-of-mass energy can be obtained, and one or two points of inflections in the dependences can be found, which may imply the phase transition from the hadronic matter to QGP. The present work is only a presentation for the methodology, but not the study of dependent relation itself.

As a thermal and statistical model, the multisource thermal model is successful in the descriptions of transverse momentum and rapidity distributions of net-baryons produced in nucleus-nucleus collisions at high energies. From these descriptions, we can extract some parameters and get some scatter plots of net-baryons in some spaces. Except for the transverse momentum and rapidity distributions, the multisource thermal model is successful in the descriptions of multiplicity distributions [13]. In addition, we can use other distributions to replace the Tsallis distribution. These distributions include, but are not limited to, the Boltzmann distribution, the Fermi-Dirac (Bose-Einstein) distribution, the Erlang distribution, and their combination of multiple components. Particularly, these multi-component distributions reflect temperature fluctuations which can be described by the Tsallis distribution which uses the entropy index to describe the degree of non-equilibrium.

More deeply, a phenomenological study on the net-baryon production in the Color Glass Condensate (CGC) formalism is performed in references [20, 24]. The transverse momentum and rapidity distributions of net-baryons are studied by the CGC formalism and related parameters are extracted [20, 24]. According to the related parameters, it is expected that the scatter plots of net-baryons in some spaces can be obtained. We notice that the maximum rapidities of net-baryons produced in central Au-Au collisions at RHIC energies obtained in the present work are slightly greater than those by the CGC formalism [20]. The shapes of the two distributions have also some differences.

In other works [25, 26] which are also in the CGC framework and based on small-coupling quantum chromodynamics (QCD), the rapidity distributions of net-baryons are studied by a gluon saturation model. It is shown that geometric scaling is exhibited in the distributions [25, 26]. We notice that the maximum rapidities of net-baryons obtained in the present work are slightly greater than those by the saturation model [25, 26], too. In addition, the shapes of the two distributions have also some differences.

**5. Conclusions**

From the above calculations and discussions, we obtain following conclusions.

(a) The transverse momentum distributions of net-baryons produced in central Au-Au collisions at 62.4 and 200 GeV are described by the Tsallis distribution which is naturally considered in the multisource thermal model. The effective temperature parameter decreases with increase of the rapidity. The entropy index does not show an obvious change trend with increase of the rapidity.

(b) The rapidity distributions of net-baryons produced in central Au-Au collisions at 62.4 and 200 GeV are studied in the framework of the multisource thermal model in which the contributions of leading target/projectile nucleon sources and target/projectile cylinders are included. Each source is described by the Tsallis statistics with the parameter values obtained from the transverse momentum distributions. The rapidity spans for the cylinders and the rapidity shifts for the leading nucleons in central Au-Au collisions at 200 GeV are greater than those at 62.4 GeV.

(c) Based on the parameter values obtained from the transverse momentum and rapidity distributions, a series of quantities are extracted by using the Monte Carlo method. As examples, the particle scatter plots at the stage of kinetic freeze-out in three types of 3D spaces are obtained. In 3D spaces $p_x$-$p_y$-$y$, $p_x$-$p_y$-$p_z$, and $\beta_x$-$\beta_y$-$\beta_z$, one can see different scatter regions of the four sources. The neighbor scatter regions overlap partly in beam direction. The particle scatter plot is a bumpy cylinder in the momentum space and a bumpy sphere in the velocity space.

(d) Particularly, the distribution range of $p_z$ is much larger than those of $p_x$ and $p_y$, and the density in small $|p_x|$ and $|p_y|$ region is greater than those in other regions. The distribution pattern in $\beta_z$ direction is very different from those in $\beta_x$ and $\beta_y$ directions, and the densities close to the maximum $|\beta_z|$ are greater than those in other regions. In 3D spaces $p_x$-$p_y$-$y$ and $p_x$-$p_y$-$p_z$, the higher center-of-mass energy is, the broader scatter plots are; and in the velocity space, the higher center-of-mass energy is, the higher densities close to the maximum $|\beta_z|$ are.

**Conflict of Interests**

The authors declare that there is no conflict of interests regarding the publication of this paper.

**Acknowledgment**

One of the authors (Fu-Hu Liu) thanks Prof. Dr. Charles Gale, Prof. Dr. Sangyong Jeon, and the members of the Physics Department of McGill University, Canada, for their hospitality, where this work was partly finished. The authors work was supported by the National Natural Science Foundation of China under Grant No. 10975095, the Open Research Subject of the Chinese Academy of Sciences Large-Scale Scientific Facility under Grant No. 2060205, and the Shanxi Scholarship Council of China.

Table 1. Values of parameters and $\chi^2/ndf$ corresponding to the fits in Figure 1, where Figures 1(a)-1(d) are for central Au-Au collisions at 62.4 GeV, and Figures 1(e)-1(h) are for central Au-Au collisions at 200 GeV. The unit of $T$ is GeV. The last three points in Figure 1(h) have not been included in the calculation of $\chi^2/ndf$ due to the low statistics.

| Figure | $y$ | $T$ | $q$ | $N_0$ | $\chi^2/ndf$ |
|---|---|---|---|---|---|
| 1(a) | 0 | $0.270\pm0.013$ | $1.023\pm0.002$ | $3.45\pm0.17$ | 0.757 |
| 1(b) | 0.65 | $0.230\pm0.012$ | $1.023\pm0.002$ | $5.13\pm0.26$ | 0.169 |
| 1(c) | 2.3 | $0.181\pm0.009$ | $1.023\pm0.002$ | $10.80\pm0.54$ | 0.546 |
| 1(d) | 3.0 | $0.158\pm0.008$ | $1.023\pm0.002$ | $14.18\pm0.71$ | 0.209 |
| 1(e) | 0 | $0.282\pm0.014$ | $1.023\pm0.002$ | $1.61\pm0.08$ | 0.764 |
| 1(f) | 0.9 | $0.239\pm0.012$ | $1.023\pm0.002$ | $1.46\pm0.07$ | 0.464 |
| 1(g) | 1.9 | $0.223\pm0.011$ | $1.023\pm0.002$ | $3.39\pm0.17$ | 0.581 |
| 1(h) | 2.9 | $0.210\pm0.011$ | $1.023\pm0.002$ | $7.79\pm0.39$ | 0.571 |

Table 2. Values of parameters and $\chi^2/ndf$ corresponding to the curves in Figs. 2(a) and 2(b), where $T$ and $q$ are taken from Table 1 for the central cylinders at $y=0$, and for the leading nucleons at $y=3.0$ (or 2.9). See ref. [14] on the discussions for the number of free parameters and the *ndf*.

| Fig. | $y_{P\max}$ | $y_{LP\min}$ | $y_{LP\max}$ | $k$ | $N_0$ | $\chi^2/ndf$ |
|---|---|---|---|---|---|---|
| 3(a) | $3.60\pm0.18$ | $1.90\pm0.10$ | $3.55\pm0.18$ | $0.21\pm0.01$ | $297\pm15$ | 0.339 |
| 3(b) | $4.35\pm0.22$ | $2.32\pm0.12$ | $4.55\pm0.23$ | $0.32\pm0.02$ | $256\pm13$ | 0.957 |

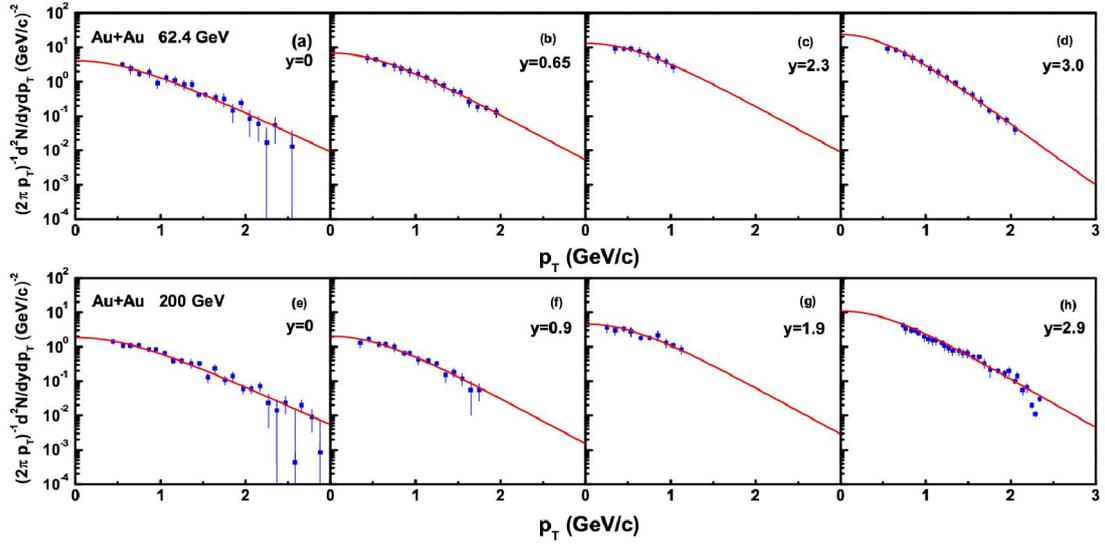

Fig. 1. Transverse momentum distributions of net-baryons produced in central Au-Au collisions at different rapidities shown in the panels. Figures 1(a)-1(d) and 1(e)-1(f) are for 62.4 and 200 GeV respectively, where the symbols represent the experimental data of the BRAHMS Collaboration [17–19] measured at the RHIC and collected together in [20], and the curves are our modeling results.

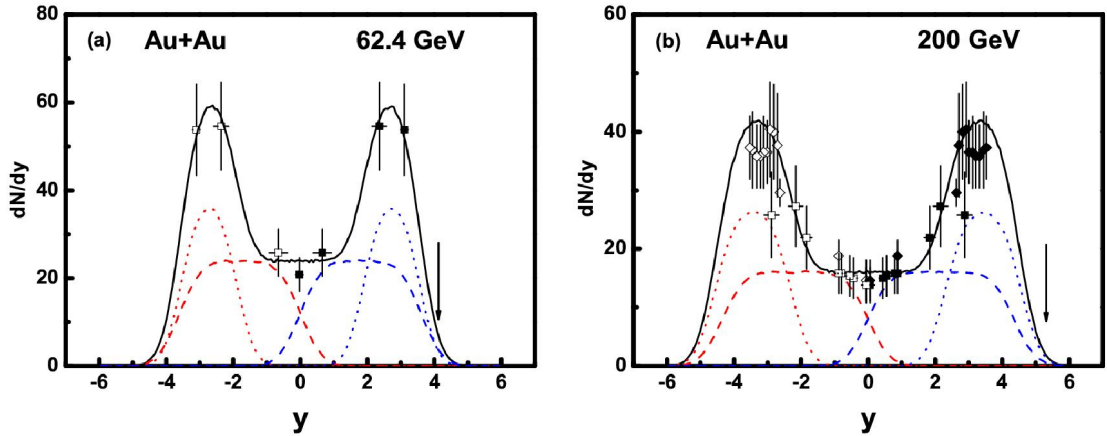

Fig. 2. Rapidity distributions of net-baryons in central Au-Au collisions at (a) 62.4 and (b) 200 GeV. The solid symbols represent the experimental data of the BRAHMS Collaboration [17–19] measured at the RHIC and collected together in [20], and the squares and diamonds are for 0–5% and 0–10% centralities respectively. The open symbols are reflected from the data around midrapidity. The curves are our calculated results, where the dotted and dashed curves from left to right are the contributions of LT, TC, PC, and LP, respectively, and the solid curves are the sum of the four sources. Arrows indicate the beam rapidities. See text on the discussion of a few net-baryons to extend beyond the beam rapidities.

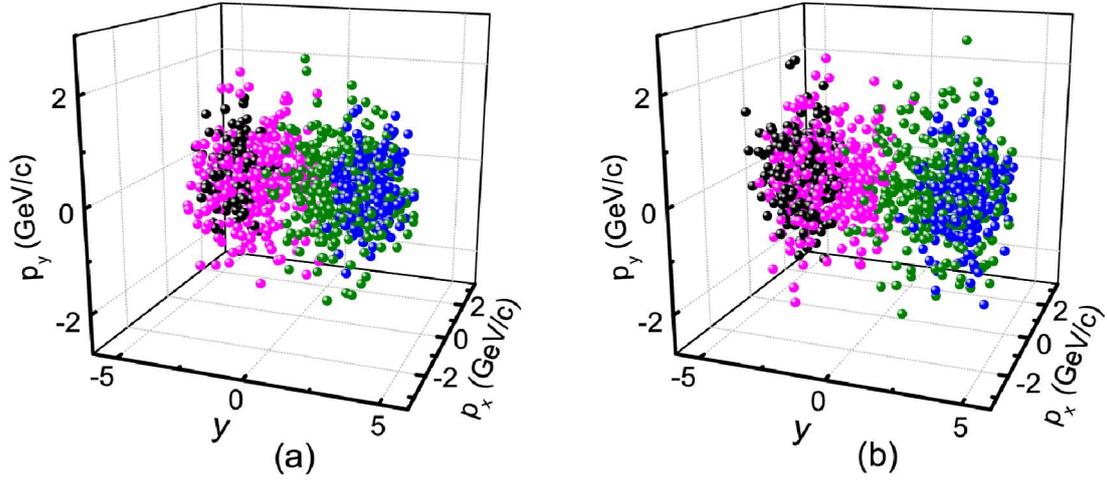

Fig. 3. Scatter plots of 1000 net-baryons in 3D $p_x$-$p_y$-$y$ space in central Au-Au collisions at (a) 62.4 and (b) 200 GeV. The black, magenta, olive, and blue balls represent orderly the contributions of LT, TC, PC, and LP from low to high in the rapidity space due to different contribution ratios.

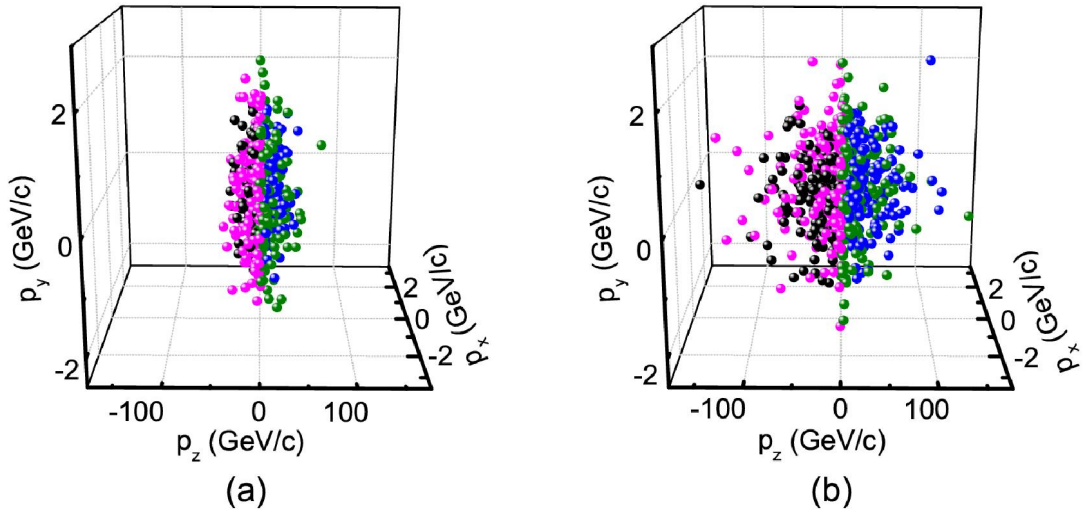

Fig. 4. The same as for Figure 3, but showing the scatter plots in 3D momentum space.

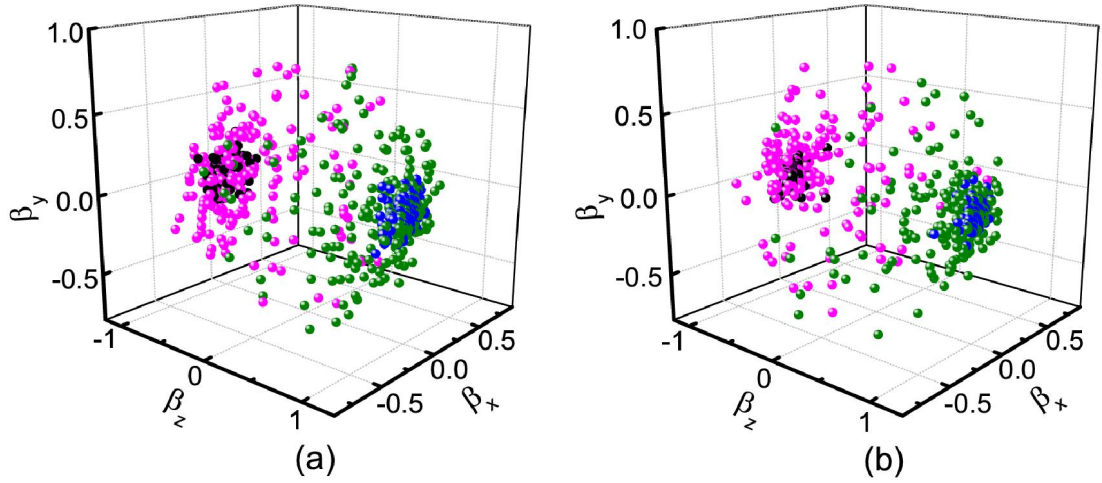

Fig. 5. The same as for Figure 3, but showing the scatter plots in 3D velocity space.